\documentclass[prl,twocolumn]{revtex4}
\usepackage{graphicx}
\usepackage{amsmath}
\usepackage{amssymb}
\usepackage{epsfig}
\usepackage{wasysym}
\usepackage{bbm}
\usepackage[colorlinks,citecolor=blue]{hyperref}
\usepackage{tabularx}
\usepackage{braket}
\usepackage{amsxtra}
\usepackage{color}

\newcommand{\Eq}[1]{Eq.~(\ref{#1})}
\newcommand{\Fig}[1]{Fig.~\ref{#1}}

\newcommand{\nn}{\nonumber\\}

\newcommand{\Rmnum}[1]{\expandafter\@slowromancap\romannumeral #1@}

\newcommand{\be}{\begin{eqnarray}}
\newcommand{\ee}{\end{eqnarray}}
\newcommand{\beq}{\begin{equation}}
\newcommand{\eeq}{\end{equation}}
\newcommand{\bpm}{\begin{pmatrix}}
\newcommand{\epm}{\end{pmatrix}}
\newcommand{\bal}{\begin{aligned}}
\newcommand{\eal}{\end{aligned}}

\newcommand{\<}{\langle}
\renewcommand{\>}{\rangle}

\newcommand{\ua}{\uparrow}
\newcommand{\da}{\downarrow}
\newcommand{\ra}{\rightarrow}

\newcommand{\e}{\epsilon}

\newcommand{\s}{{\sigma}}

\newcommand \ti[1]{}

\begin{document}

\title{Symmetry protected topological Luttinger liquids and the phase transition between them
}
\author{Hong-Chen Jiang$^1$, Zi-Xiang Li$^2$, Alexander Seidel$^3$ and Dung-Hai Lee$^{4,5}$\footnote{Corresponding author}}
\affiliation{$^1$ Stanford Institute for Materials and Energy Science, SLAC and Stanford University, 2575 Sand Hill Road, Menlo Park, CA 94025, USA.\\ $^2$ Institute for Advanced Study, Tsinghua University, Beijing 100084, China.\\$^3$ Department of Physics, Washington University, St. Louis, MO 63130, USA.\\$^{4}$ Department of Physics, University of California, Berkeley, CA 94720, USA.\\$^5$ Materials Sciences Division, Lawrence Berkeley National Laboratory, Berkeley, CA 94720, USA.}

\date{\today}
\begin{abstract}
We show that a doped spin-1/2 ladder with antiferromagnetic intra-chain and ferromagnetic inter-chain coupling 
is a symmetry protected topologically non-trivial Luttinger liquid. Turning on a large easy-plane spin anisotropy drives the system to a topologically-trivial Luttinger liquid. 
Both phases have full spin gaps and exhibit power-law superconducting pair correlation. The Cooper pair symmetry is singlet $d_{xy}$ in the non-trivial phase and triplet $S_z=0$ in the trivial phase. The topologically non-trivial Luttinger liquid exhibits gapless spin excitations in the presence of a boundary, and it has no non-interacting or mean-field theory analog even when the fluctuating phase in the charge sector is pinned.  As a function of the strength of spin anisotropy 
there is a topological phase transition upon which the spin gap closes.   We speculate these Luttinger liquids are relevant to the superconductivity in metalized  integer spin ladders or chains.
\end{abstract}

\pacs{64.70.Tg, 05.30.Rt}
\maketitle

Symmetry protected topological states (SPTs)\cite{Schnyder2008,Kitaev2009,Chen2011} have attracted lots of interest in condensed matter physics recently. These states  do not break any symmetry and are fully gapped, except at the boundary. The gapless boundary excitations of SPTs are protected by symmetry, hence are very different from ``accidental boundary states'' found at, say, semiconductor surfaces (due to ``dangling bonds'').  Searching for real materials realizing SPTs is a very active area of research.

The best known examples of SPT are topological insulators and superconductors\cite{Schnyder2008,Kitaev2009}. In addition to these there are ``bosonic'' SPTs\cite{Chen2011}. A classic example is the antiferromagnetic (AF) spin-1 Heisenberg chain\cite{haldane,aff}, which is gapped in the bulk but possesses gapless spin-1/2 excitations on the boundary\cite{kennedy}. These spin-1/2 boundary excitations are protected so long as the SO(3) spin rotation symmetry is respected (in fact even its Z$_2\times$ Z$_2$ subgroup is sufficient for the protection)\cite{pollman,Chen1D}. Unlike topological insulators and superconductors, which can be realized in non-interacting (or mean-field) theories, the  novel collective mode dynamics of spin-1 chains is caused by strong interaction.  So long as the protective symmetry is unbroken, inequivalent SPTs are connected by topological phase transitions where the bulk gap closes. 

In this paper we demonstrate that in one space dimension there exists topologically inequivalent (gapless) Luttinger liquids whose distinction is also protected by symmetry. 
These Luttinger liquids exhibit power-law Cooper pair correlation, hence are phase fluctuating superconductors. Interestingly the pairing symmetries are different in topologically inequivalent phases. Because the spin gap in the topological non-trivial Luttinger liquid is due to collective mode dynamics, even after pinning the fluctuating phase in the charge sector it has no non-interacting or mean-field theory analog, hence is different from that discussed in Ref.\cite{fid,sau,keselman}. They are also very different from the Weyl\cite{Weyl} or Dirac\cite{Dirac} semi-metals whose existence is not protected by a on-site (non-crystal) symmetry. We also study the phase transition between these Luttinger liquids and show that the quantum critical point exhibits central charge $c=2$. 

We begin by briefly reviewing a known topological phase transition between different phases of a spin-1 chain. The Hamiltonian under consideration is\cite{hida}
\vspace{-0.2in}
\be
H=J\sum_{i=1}^L \vec{S}_i\cdot\vec{S}_{i+1}+D\sum_{i=1}^L S_{i,z}^2.
\label{sp1H}\ee
In \Eq{sp1H} $i$ labels the sites of a one-dimensional lattice under periodic boundary condition (PBC) (i.e., $L+1\equiv 1$)   and $\vec{S}$ is the spin-1 operator. The last term breaks the SO(3) spin rotation symmetry down to U(1)$\times$ Z$_2$. It favors $\<S_{i,z}^2\>$ to be zero (non-zero) for $D>0$ ($D<0$).   For $D=0$ the ground state of \Eq{sp1H} is a non-trivial SPT (the Haldane phase)\cite{haldane,aff}. On the other hand, for $D/J>>1$ the ground state is a product state with $S_z=0$ on every site. Both phases have a full spin gap and do not break the U(1)$\times$ Z$_2$ symmetry, but the Haldane phase possesses gapless boundary excitations while the ``large D'' phase does not\cite{hida}. As a function of $D$ there is a topological phase transition occurring around $D/J\approx 1$. The central charge of the critical theory is estimated to be 1\cite{hida}.

\begin{figure}[htp]
\includegraphics[width=8.6 cm,height=4.3 cm]{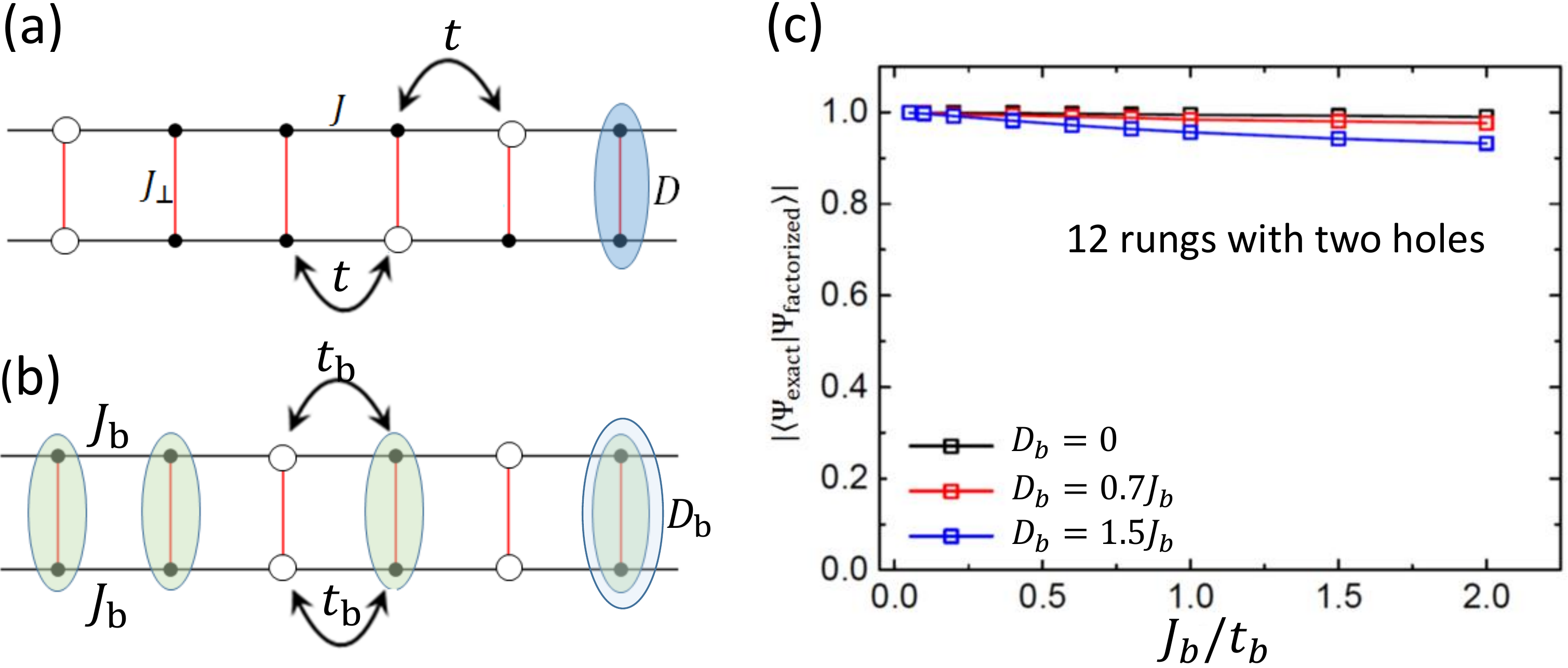}
\caption{(Color online) (a) A caricature of \Eq{tj}. The red  vertical bonds are ferromagnetic  while the black horizontal bonds are AF.
(b) A caricature of \Eq{Hb}. The triplet bosons are denoted by the light green ellipses. When doped these  bosons can hop to nearest neighbor empty rungs. Each triplet boson experiences a spin anisotropy term and interacts with neighboring bosons via AF exchange interaction.  (c) The overlap between the exact ground state of 11 triplet bosons on a 12 rung lattice and the factorized wavefunction, as a function of $J_{\sf b}/t_{\sf b}$ for several value of $D_{\sf b} /J_{\sf b}$}
\label{fig1}
\end{figure}

Now we consider doping the above spin chain. To model doping we construct the following
``t-J'' type ladder Hamiltonian (see \Fig{fig1}(a)):
\be
&&H=-t\sum_{i=1}^L\sum_{\alpha=1}^2\sum_{\s=\uparrow,\downarrow}\left(c^\dagger_{\alpha,i+1,\s}c_{\alpha,i,\s}+h.c.\right)\nn&&+J\sum_{i=1}^L\sum_{\alpha=1}^2 \vec{S}_{\alpha,i}\cdot\vec{S}_{\alpha,i+1}+J_\perp\sum_{i=1}^L\vec{S}_{1,i}\cdot\vec{S}_{2,i}\nn&&
+D\sum_{i=1}^L\left(S_{1,i,z}+S_{2,i,z}\right)^2.\label{tj}\ee
Here $\alpha=1,2$ labels the two chains and note that we only allow intra-chain hopping. This allows us to interpret $\alpha$ as labeling two different orbitals of an atom later, as atomic orbitals do not hybridize. In general as long as the strength of the inter-chain hopping is
weak compared with $|J_\perp|$ we do not expect any qualitative change in the results. \Eq{tj} is supplemented with the Hilbert space constraint that there is at most one electron per site (i.e. $n_{\alpha,i}\le 1$). \Eq{sp1H} can be shown to be the {\it effective} Hamiltonian at half filling (one electron per site) 
under the conditions $t=0$, $J_\perp<0$ (FM) and $|J_\perp|>>J,D$.

Now consider doping holes into the system. We first consider large $|J_\perp|$ so that minimizing the rung exchange energy requires doped holes to form ``vertical  pairs''. Under such condition the two electrons on each rung are effectively a spin-triplet boson which can hop into the empty rung created by doping  via a second order virtual process $t_{\rm b}={\cal O}(t^2/|J_\perp|)$ (see \Fig{fig1}(b)). The resulting
effective Hamiltonian is given by
\be\label{Hb}
H_{\sf b}&&=-t_{\sf b} \sum_{i=1}^L\sum_{m=-1}^1 b^\dagger_{i,m}b_{i+1,m}+h.c.+J_{\sf b} \sum_{i=1}^L {\vec S}_i\cdot {\vec S}_{i+1}\nn&&+D_b\sum_{i=1}^L S_{i,z}^2\,,
\ee
where $m=-1,0,1$ labels the three states of  spin 1, ${\vec S}_i$ is the total spin operator of the i$^{\rm th}$ rung, $J_{\sf b}={\cal O}(J) , D_{\sf b}={\cal O}(D)$ and both are positive.

To illustrate the  basic physics it is instructive to consider the limit $t_{\sf b} >> J_{\sf b}$ and $t_{\sf b} >> D_{\sf b}$. Under such condition, one can generalize the result of \cite{Ogata1, Parola, Ogata2} and show that the ground state wave function   exhibits spin-charge separation, namely,
\be\label{wf}
&&\<0|b_{x_1,m_1}\dotsc b_{x_N,m_N}|\psi_0\>= f(x_1\dotsc x_N)g(m_1\dotsc m_N)\nn
&&{\rm for~} x_1<\dotsc<x_N<x_1+L.
\ee
In \Eq{wf}  $f(x_1\dotsc x_N)$ is the charge and $g(m_1\dotsc m_N)$ is the spin wavefunction, both are subjected to PBC. $f(x_1\dotsc x_N)$ is the ground state of {\em spinless} hard-core bosons with nearest neighbor hopping.
$g(m_1\dotsc m_N)$ is the ground state of an effective spin Hamiltonian defined on the ``squeezed'' lattice, i.e. the lattice formed by deleting the holes. The effective spin Hamiltonian has the same form as the last two terms of \Eq{Hb} except the parameters are renormalized: 
$J_{\sf eff}=J_{\sf b}\braket{\hat n_i \hat n_{i+1}}_f\times L/N>0$ 
and $D_{\sf eff}=D_b\<n_i\>_f\times L/N=D_b>0$, where $\hat n_i$ is the boson number operator and $\<..\>_f$ denotes the expectation value computed using the hard-core boson wave function $f$. In the case of $D_{\sf b}=0$, $g$ is the ground state of the spin-1 AF Heisenberg chain\cite{haldane,aff}. For  $D_{\sf b}>>J_{\sf b}$, $g$ is the ground state wavefunction of the large-$D$ phase of \Eq{sp1H}\cite{hida}.

As a preliminary check of the validity of the factorized wavefunction we show in \Fig{fig1}(c) the overlap of the exact ground state and the factorized wavefunction for a 12-site spin-1 chain doped with one bosonic hole.
Although the factorized wavefunction only becomes exact in the  $J_{\sf b}/t_{\sf b}\ra 0$ and $D_{\sf b}/t_{\sf b}\ra 0$ limit, the computed overlap is close to unity for
a substantial range of $J_{\sf b}/t_{\sf b}$ and  $D_{\sf b}/t_{\sf b}$.
Due to spin-charge separation while both the $D_b=0$ and the large $D_b$ phases are described by the same gapless Luttinger liquid in the charge sector, they differ topologically in the spin sector. 
Thus we have two topologically inequivalent Luttinger liquids !

In order to check whether the above Luttinger liquids survive when the parameters of \Eq{tj} move away from the limit considered above,  and to study the phase transition between these Luttinger liquids,  we performed large scale density-matrix renormalization group\cite{DMRGRef} (DMRG) calculations. Specifically we studied the ground state and spin excitations of \Eq{tj} for  $t=1,J=0.3, J_\perp=\pm 3$ as a function of $D$ at $5\%$ hole doping.  We perform up to 24 sweeps and keep up to $m$=10000 states in each DMRG block with a typical truncation error $\epsilon\sim10^{-8}$ for PBC and $\epsilon\sim 10^{-11}$ for {\it open} boundary conditions (OBC). This leads to excellent convergence for the results that we report here. For more details we refer to the supplementary material A.  

\begin{figure}[htp]
\includegraphics[width=8.6 cm,height=4.4 cm]{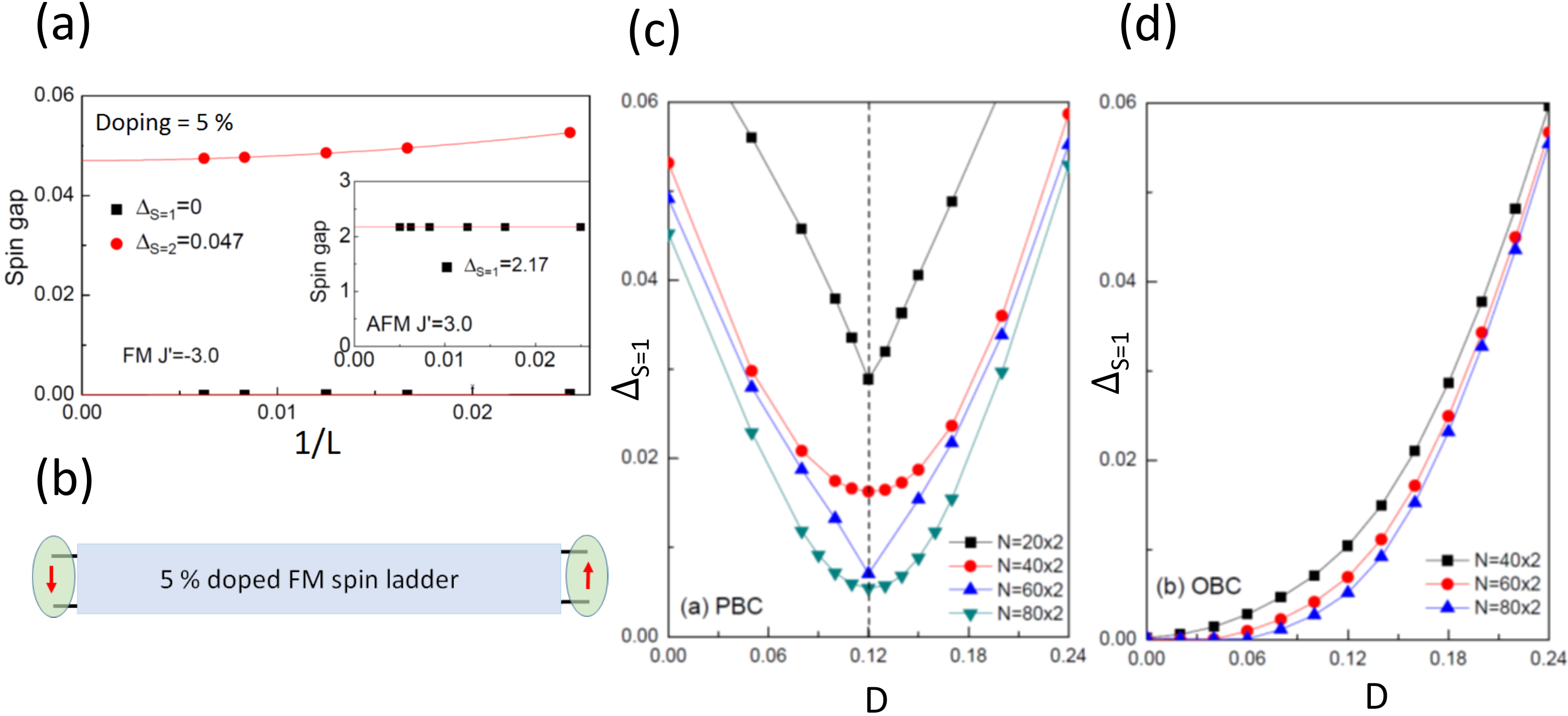}
\caption{(Color online) (a) The spin gap $\Delta_S$ at $D=0$ with $5\%$ hole doping for different ladder length $L$. 
The black (red) symbols mark the spin-1(2) excitation gap. The main panel is for FM rung coupling and the inset shows the spin 1  gap for AF rung coupling. (b) A caricature for the spin gapped and charge gapless bulk with spin-1/2 boundary excitations (red arrows).(c) The spin 1 gap as a function of D under PBC. (d) The spin-1 gap as a function of D under OBC.
}
\label{fig2}
\end{figure}

First let us focus on $D=0$. The main panel of \Fig{fig2}(a) shows the values of the spin-1 and spin-2 excitation  gaps ($\Delta_{S=1}, \Delta_{S=2}$) under OBC for FM rung coupling and several ladder length. Here the spin gap is defined as $\Delta_S=E_0(S_z=S)-E_0(S_z=0)$, where $E_0(S_z)$ is the ground state energy of the system with total spin $S_z$. Interestingly, while $\Delta_{\rm S=2}$ (red circles) is {\it non-zero},  $\Delta_{\rm S=1}$ (black squares) is {\it zero}. This is paradoxial if one takes $\Delta_{\rm S=1}=0$ as implying the presence of gapless bulk $S=1$ excitations, as  spin-2 excitations can be made up from two spin 1 excitations. 
However, this behavior is totally consistent with the presence of gapless boundary spin-1/2 excitations (\Fig{fig2}(b)).
Two spin-1/2s make up a spin-1 and a spin-0 hence will not contribute to the spin 2 excitation. In \Fig{fig2}(c) and (d) we compare the spin gap under PBC and  OBC. At $D=0$ $\Delta_{S=1}$ is non-zero under PBC but is zero under OBC. This directly confirms the presence of gapless boundary excitations.
For comparison in the inset of \Fig{fig2}(a) we show the spin-1 gap when the rung coupling is AF ($J_\perp=+3$). As expected, the spin 1 excitation is gapped for OBC (and PBC).

\begin{figure}[htp]
\includegraphics[scale=0.48]{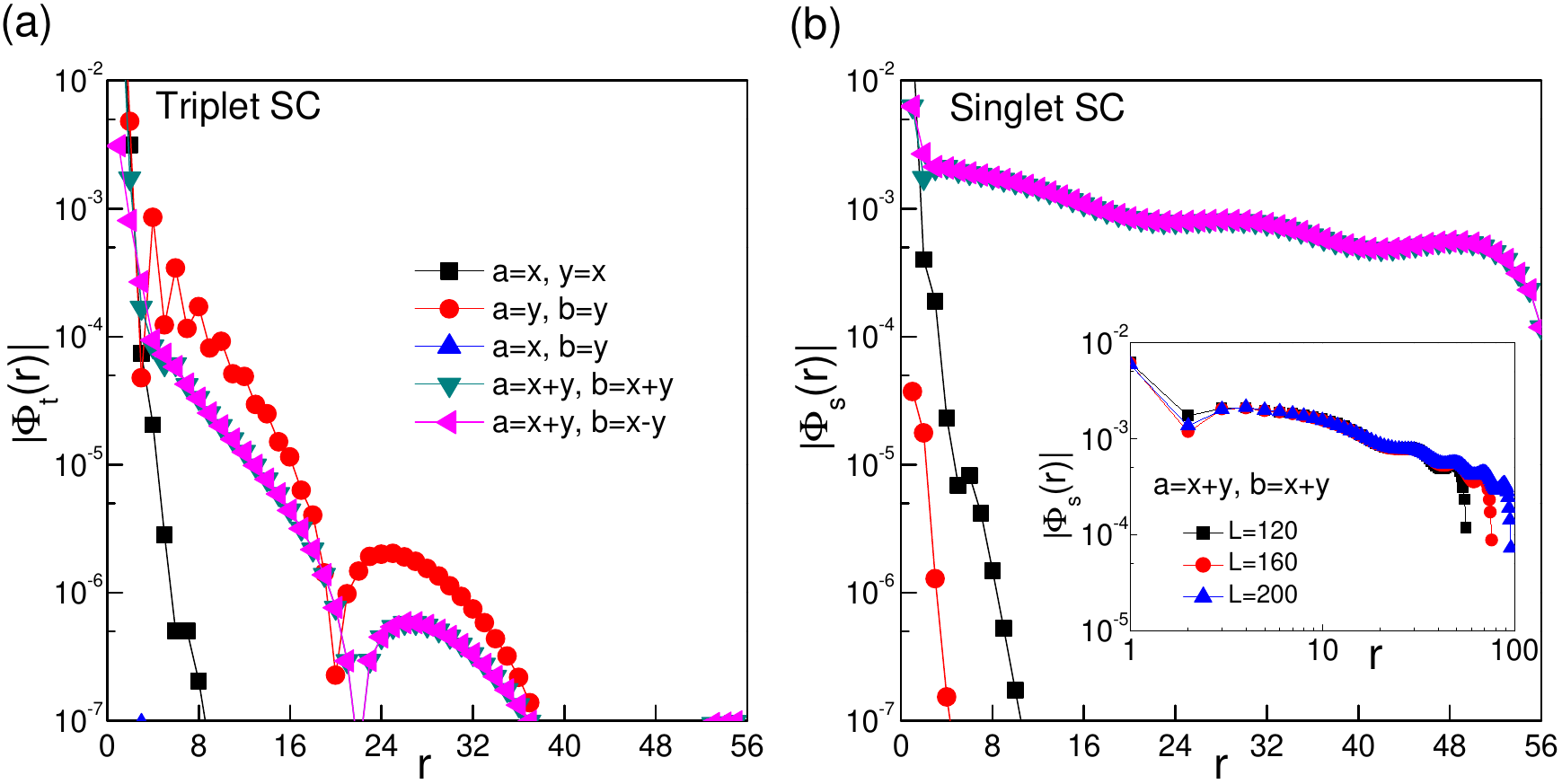}
\caption{(Color online) The triplet (a) and singlet(b)  Cooper pair correlation functions at $D=0$. See main text for detailed explanation. The ladder size is $120\times 2$. The inset shows the singlet Cooper pair correlation functions for $L$=120, 160 and 200 in the ln-ln scale.
}
\label{fig3}
\end{figure}

\Fig{fig3}(a) and (b) show the {\it absolute value} of the Cooper pair correlation function
\be
&&\Phi_{s(t),ab}(r)=\<\Delta^\dagger_{s(t),a}(i)\Delta_{s(t),b}(i+r)\> {\rm~~where}\nn
&&\Delta^\dagger_{s(t),a}(i)=\left(c^\dagger_{i\ua}c^\dagger_{i+a\da}-(+)c^\dagger_{i\da}c^\dagger_{i+a\ua}\right)/\sqrt{2}.
\label{pair}\ee
In \Eq{pair} $i$ sits on the lower chain and $a=x,y, (x\pm y)$ denotes the nearest neighbor (next nearest neighbor) along the chain, rung (diagonal) directions respectively.  The symbols $s$ and $t$ stand for  singlet and triplet.
The triplet pair correlation, panel (a), exhibits exponential decay. On the surface this is counter-intuitive because the rung coupling is strongly FM. Upon further reflection this is  expected because the spin sector is in the Haldane phase, hence all triplet spin correlation should decay exponentially. \Fig{fig3}(b) shows the much slower decaying singlet pair correlation. The pairing channel which exhibits the strongest  correlation (consistent with power-law decay, see inset in \Fig{fig3}(b)) is associated with $a=x\pm y$ and $b=x\pm y$. The overlapping cyan and pink symbols in \Fig{fig3}(b) denote $(a,b)=(x+y,x+y)$ and $(a,b)=(x+y,x-y)$, respectively. The degeneracy is caused by taking the absolute value. The  actual correlations  have opposite sign. Such sign structure can be interpreted as $d_{xy}$-pairing on the ladder.

If  we view the $\alpha=1,2$ as labeling two orbitals of an atom and use $c^\dagger_{1,i,\s}$ and $c^\dagger_{2,i,\s}$ to denote the corresponding electron creation operator, 
then the Cooper pair which exhibits power-law correlation is created by
\be
(c^\dagger_{1,i\ua}c^\dagger_{2,i+1\da}-c^\dagger_{1,i\da}c^\dagger_{2,i+1\ua})-(c^\dagger_{2,i\ua}c^\dagger_{1,i+1\da}-c^\dagger_{2,i\da}c^\dagger_{1,i+1\ua}).\nonumber\ee
This is {\it spin singlet, orbital antisymmetric and odd parity} pairing. 
Our result suggests the  doped spin-1 ladder studied above is a spin-gapped charge-gapless  liquid with power-law superconducting correlations.

Now let's increase the value of $D$. In \Fig{fig2}(c) we show the spin gap under PBC as a function of $D$. The gap reaches a minimum at $D_c\approx 0.12$ and the  value decreases with increasing ladder length consistent with a zero spin gap at $D_c$. A closer inspection of \Fig{fig2}(c) shows an even-odd effect in terms of the number of  hole pairs. We explain such effect in the supplementary material B. 
In \Fig{fig2}(d) we plot the spin-1 gap under OBC as a function of $D$ for several values of $L$ at $5\%$ doping. The result is consistent with the absence of a spin gap for $D<D_c$ while a finite spin gap remains for $D>D_c$ as $L\ra\infty$. The results of \Fig{fig2}(a,c,d)
support the statement that the for $D<D_c$ the spin sector realizes a non-trivial SPT (the Haldane phase) with gapless edge excitation, while for $D>D_c$ the spin state is a trivial SPT (the large $D$ phase) with no gapless edge excitations. A topological phase transition occurs at $D_c$ where the spin gap closes.
\begin{figure}[htp]
\includegraphics[width=8.6 cm, height=4.4 cm]{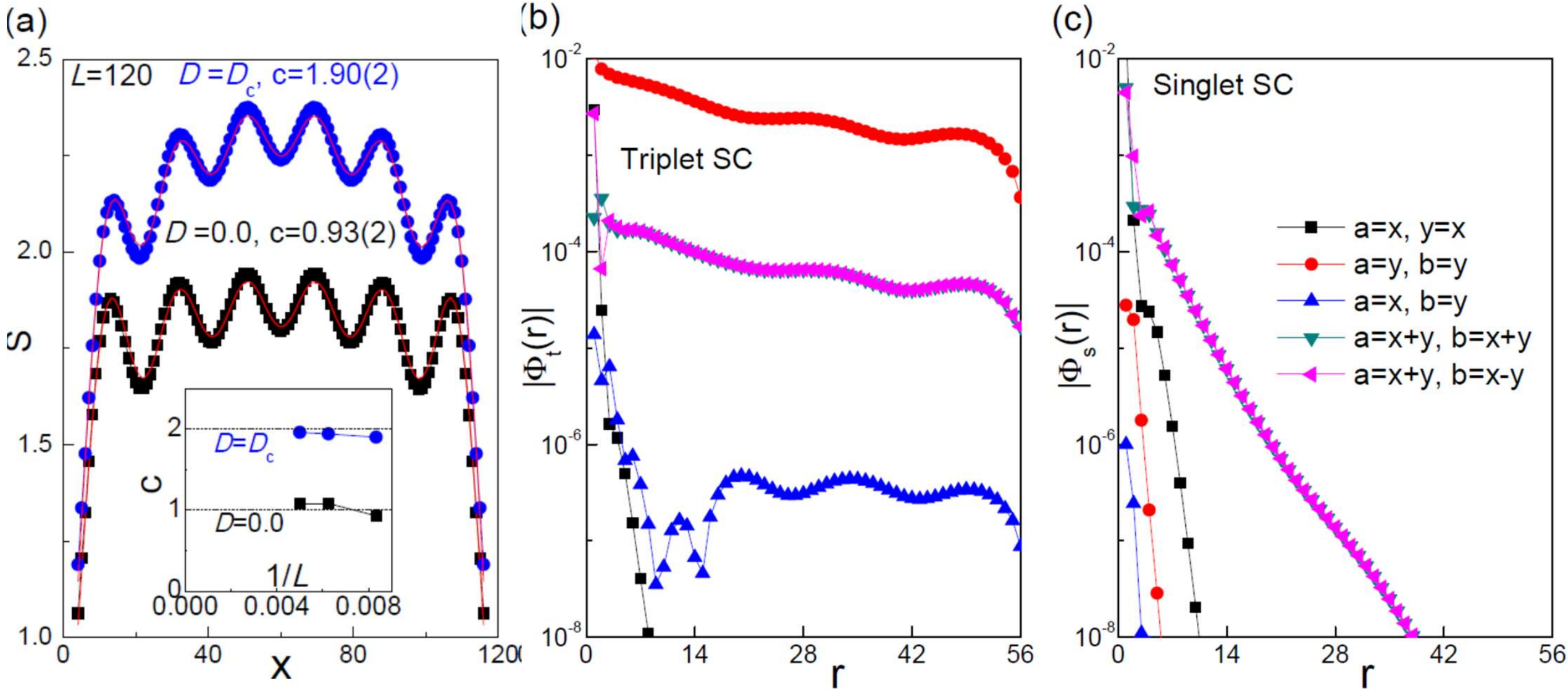}
\caption{(Color online) (a) The entanglement entropy as a function of number of rungs in an untraced subsystem for $D=0$ and  $D=D_c$ under OBC. The length of the ladder is $L=120$. The red line is the fit using Eq.(70) in Ref.\cite{osci}. The inset shows the deduced central charge as a function of $1/L$ for $L=120,160,200$.  The triplet (b) and singlet (c) Cooper pair correlation functions at $D=0.2>D_c$. The ladder size is $120\times 2$ and the doping level is $5\%$.
}
\label{fig4}
\end{figure}

In \Fig{fig4}(a) we plot the entanglement entropy as a function of
number of rungs in an untraced subsystem  for OBC. The oscillatory behavior is a consequence of the Friedel-like oscillation induced by the boundary effect due to the finite density wave susceptibility associated with hole pairs. This is common in 1D Luttinger liquids. Using the data of the longest ladder ($L=200$) the best fit using Eq.(70) in Ref.\cite{osci} gives $c\approx 1.08$ (consistent with $c=1$) for $D=0$ and $c\approx 1.96$ (consistent with $c=2$) for $D=D_c$. The central charge at $D_c$ is consistent with the sum of the central charge  associated with the gapless charge sector (namely $c_{\rm charge}=1$) and that of the spin sector (namely $c_{\rm spin}=1$) at the  topological phase transition point.

Next we study the superconducting pair correlation in the large $D$ phase.  In \Fig{fig4}(b,c) we show the absolute value of the spin triplet and spin singlet Cooper pair correlation function. In this case all singlet pair correlations decay exponentially.  In contrast the tripet correlation decays much slower (consistent with power-law decay). In part (b) the cyan and pink symbols overlap signifying the same absolute value. The actual values have opposite sign. The spin state of the triplet Cooper pair that show the strongest correlation is $S=1, S_z=0$. Again if we interpret the two chains as corresponding to the two atomic orbitals 1,2 the triplet Cooper pair in question is created by
 \be
&&(c^\dagger_{1,i\ua}c^\dagger_{2,i\da}+c^\dagger_{1,i\da}c^\dagger_{2,i\ua})+A(c^\dagger_{1,i\ua}
c^\dagger_{2,i+1\da}+c^\dagger_{1,i\da}c^\dagger_{2,i+1\ua})\nn&&-A(c^\dagger_{2,i\ua}
c^\dagger_{1,i+1\da}+c^\dagger_{2,i\da}c^\dagger_{1,i+1\ua}),
\label{trc}\ee where $A$ is a constant. This is {\it spin triplet, orbital anti-symmetric and even parity} Cooper pairing. 

The topologically non-trivial and trivial Luttinger liquids described so far can also be obtained from bosonization. However, for space consideration we refer to the supplementary material C.

Doped spin ladders with AF rung coupling have been extensively studied (see Ref.[3-7] of supplementary material). At half-filling the spin state is a good example of the resonant-valence-bond state\cite{Anderson}. Upon doping the system becomes a spin gapped Luttinger liquid with $d_{x^2-y^2}$ Cooper pairing. In this paper we studied doped spin ladders with FM rung coupling. At half-filling the spin state can be either a non-trivial SPT (Haldane phase) or a trivial one (the large $D$ phase). Upon doping they also become phase fluctuating superconductors (
spin gapped Luttinger liquids) with different Cooper pair symmetries. This suggests the possibility of  superconductivity in metalized integer spin ladders or chains. A particular interesting question is the relevance of our work to the superconductivity in FeSe. FeSe becomes nematic below 90K while maintains paramagnetic down to the lowest measured temperature. Superconductivity onsets at $\sim 8K$ 
with $\Delta/\e_F={\cal O}(1)$  suggestive of real space pairing. In Ref.\cite{wkl,S1AFMH} it is proposed that the nematic, yet paramgnetic, state is due to the spontaneous formation of spin-1 chains.  If so the superconducting pairing considered here could be relevant.

{{\bf Acknowledgement} We thank Zheng-Yu Weng for communicating their upcoming publication on two holes in a similar t-J ladder\cite{Weng2017} to us.  This Work was primarily funded by the U.S. Department of Energy (DOE), Office of Science, Office of Basic Energy Sciences (BES), Materials Sciences and Engineering Division under Contract no DE-AC02-05CH11231 within the Theory of Materials Program (KC 2301). 
Work by HCJ was supported by the U.S. Department of Energy (DOE), Office of Science, Office of Basic Energy Sciences, Division of Materials Sciences and Engineering, under Contract No. DE-AC02-76SF00515. }

\bibliographystyle{ieeetr}
\bibliography{bibs}
\end{document}


\title{Supplementary material for symmetry protected topological Luttinger liquids}
\author{Hong-Chen Jiang$^1$, Zi-Xiang Li$^2$, Alexander Seidel$^3$ and Dung-Hai Lee$^{4,5}$\footnote{Corresponding author}}
\affiliation{$^1$ Stanford Institute for Materials and Energy Science, SLAC and Stanford University, 2575 Sand Hill Road, Menlo Park, CA 94025, USA.\\ $^2$ Institute for Advanced Study, Tsinghua University, Beijing 100084, China.\\$^3$ Department of Physics, Washington University, St. Louis, MO 63130, USA.\\$^{4}$ Department of Physics, University of California, Berkeley, CA 94720, USA.\\$^5$ Materials Sciences Division, Lawrence Berkeley National Laboratory, Berkeley, CA 94720, USA.}

\date{\today}
\begin{abstract}

\end{abstract}

\pacs{64.70.Tg, 05.30.Rt}
\maketitle
\section{A: Some details of the DMRG calculations}

We determine the ground state phase diagram and properties of the model Hamiltonian in Eq. (2) by extensive and highly accurate DMRG \cite{DMRGRef} simulations. For the present study, we consider both PBC and OBC, and keep up to $m = 10000$ number of states in each DMRG block with up to 24 sweeps to get converged results. The typical truncation error is of the order $\epsilon\sim 10^{-8}$ for PBC and $\epsilon\sim 10^{-11}$ for OBC. This allows us to get accurate results for both systems including ground state energy and entanglement entropy.

 For the critical theory in one dimension with OBC, the central charge of the conformal field theory can easily be extracted by fitting the von Neumann entanglement entropy to the analytical form \cite{OSCI}
 \begin{eqnarray}\label{Eq:EEFit}
 S(x) &=& \frac{c}{6}\ln\left [ \frac{4(L+1)}{\pi} \sin\left(\frac{\pi(2x+1)}{2(L+1)} \right) \right] \\%
 &-&\frac{\sin(k_F^\prime (2x+1)}{|\sin k_F^\prime| \frac{4(L+1)}{\pi} \sin\left[ \frac{\pi(2x+1)}{2(L+1)} \right]} + {\rm const}, \nonumber
 \end{eqnarray}
 where $L$ is the length of system and $x$ is the number of rungs in the untraced subsystem. $c$ is the central charge and $k_F^\prime$ is a fitting parameter. Performing the fit of $S(x)$ using Eq.(\ref{Eq:EEFit}) to the data in Fig.4(a) with different system sizes, we get the central charge $c\approx 1$ for $D=0$ and $c\approx 2$ for $D=D_c$.

\section{B: The even-odd effect in Fig.2(c) of the main text}
To understand this effect, we point out that the same even-odd effect should exist at half-filling as a function of the number of rungs (Eq.2 of the main text) or the number of sites (Eq.3 of the main text). This is because $Q=\prod_{i=1}^L e^{i\pi S_i^x}$ (where $S_i^x$ is the total x-component spin operator on the i$^{\rm th}$ rung (site)) commutes with the Hamiltonian, hence is a good quantum number. It is straightforward to show that for the Haldane phase $Q=1$ while for the large $D$ phase $Q=(-1)^L$.
Therefore, for odd $L$ a $Q=-1$ excited state crosses the $Q=1$ Haldane state at  $D_c$. On the other hand, for even $L$ such level crossing is avoided. This presence/absence of avoided crossing is responsible for the even-odd effect in $\Delta_{S=1}$. For the doped case the number of rungs (sites) in the ``squeezed'' spin ladder (chain) depends on the number of hole pairs (holes) which explains the even-odd effect in Fig.2(c) of the main text.

\section{C: Bosonization}
 In this appendix we look at the problem discussed in the main text from the bosonization point of view. Fermionic Hubbard ladder has been studied  by bosonization extensively in previous works \cite{Balents,Lin,Tsvelik-2005,Tsvelik-2011,naoto}. Our results are consistent with that reported in Ref.\cite{naoto,Tsvelik-2011}. However, the physics of doped two topologically distinct phases has not been discussed before.
Our purpose is to show that the topologically inequivalent Luttinger liquids discussed in the main text are indeed stable low energy phases.

Our starting point is the following fermion Hubbard ladder with ferromagnetic (FM) rung coupling and  single-ion spin anisotropy terms. The intra-chain antiferromagnetic (AF) exchange will be generated by the superexchange and for sufficiently large U charge fluctuation will be suppressed.
\bea\label{Hamiltonian}
H &=& H_0 + H_{\rm int} \nonumber\\
H_0 &=& -t \sum_{i}\sum_{\alpha=1}^2\sum_{\s=\ua,\da}(c^\dagger_{i\alpha\sigma}c_{i+1\alpha\sigma} + h.c)  \nonumber\\
H_{\rm int} &=& U\sum_{i}\sum_{\alpha=1,2}(n_{i\alpha\uparrow} - \frac{1}{2})(n_{i\alpha\downarrow} - \frac{1}{2}) \nonumber\\
 &+& J_\perp \sum_i \vec{S}_{i,1} \cdot \vec{S}_{i,2} + D \sum_i(S_{i,1,z} + S_{i,2,z})^2
\eea
Here $\alpha=1,2$ labels two chains and $\sigma=\uparrow,\downarrow$ denotes spin polarization. $\vec{S}_{i,\alpha}$ represents spin operator on $\text{i}^{\text{th}}$ site of chain-$\alpha$, given by $$\vec{S}_{i,\alpha}=\frac{1}{2} \sum_{s,s^\prime=\ua,\da}c^\dagger_{i\alpha,s} \vec{\sigma}_{s s^\prime} c_{i\alpha s^\prime}\,.$$ The parameter $J_\perp$ is inter-chain spin exchange, it is FM if $J_\perp<0$; $D$ represents the  single-ion anisotropy. It is easy-plane if $D>0$.

The $H_{\rm int}$ term in \Eq{Hamiltonian} can be rewritten as
\bea
H_{\rm int} &=& U^\prime\sum_{i,\alpha=1,2}(n_{i\alpha\uparrow} - \frac{1}{2})(n_{i\alpha\downarrow} - \frac{1}{2})
 + J_{\perp,z} \sum_i S^z_{i,1}S^z_{i,2}\nonumber\\ &+& J_{\perp,xy} \sum_i (S^x_{i,1}S^x_{i,2}+S^y_{i,1}S^y_{i,2})
 \label{hint}
\eea
Here $J_{\perp,z}=J_\perp+2D$, $J_{\perp,xy} = J_\perp$ and $U' = U-D/2$. Thus for non-zero $D$, the inter-chain FM coupling becomes anisotropic.


In the following we apply Abelian bosonization method to study \Eq{Hamiltonian} and \Eq{hint}. The low energy fermion operators which annihilate the left and right moving fermions near the Fermi points of
$H_0$ in \Eq{Hamiltonian} are rewritten in terms bosonic variables as follows
\bea
\psi_{R(L),\alpha,\sigma}(x) = \frac{U_{R(L),\alpha,\sigma}}{\sqrt{2\pi a}} :e^{ i(\phi_{\alpha,\sigma}(x) \pm \theta_{\alpha,\sigma}(x)) }:
\eea
where $\alpha$ and $\sigma$ are the chain and spin indices as before. $a$ is a short-distance cutoff. The ``branch'' index $R(L)$ denotes right(left) moving fermions, respectively. $U_{R(L),\alpha,\sigma}$ is a Klein factor which ensures the anti-commutation relation between fermions with different chain, spin and branch indices. For convenience, we perform linear transformations on the four $\phi_{\alpha,\sigma}$ and $\theta_{\alpha,\sigma}$ modes to get the four symmetric/anti-symmetric (+/-), charge/spin (c/$\s$),  modes:
\bea
\phi_{\pm,c} &=& \frac{1}{2}\left[(\phi_{1,\uparrow} \pm  \phi_{2,\uparrow}) + (\phi_{1,\downarrow} \pm  \phi_{2,\downarrow})\right] \nonumber\\
\phi_{\pm,\s} &=& \frac{1}{2}\left[(\phi_{1,\uparrow} \pm  \phi_{2,\uparrow}) - (\phi_{1,\downarrow} \pm  \phi_{2,\downarrow})\right].
\eea
In the following, we skip the details of the analysis and just summarize the main results.

The bosonized kinetic energy Hamiltonian is given by
\bea
H_0 = \int dx \sum_{\substack{\mu=\pm, \\ \beta=c,\s}} \frac{v_{\mu,\beta}}{2\pi}\left[ K_{\mu,\beta} \Pi^2_{\mu,\beta}(x) + \frac{1}{K_{\mu,\beta}}(\partial_x \phi_{\mu,\beta})^2\right] \nonumber\\
\eea
where $v_{\mu,\beta}$ and $K_{\mu,\beta}$ are Fermi velocities and Luttinger parameters renormalized by the Hubbard and spin exchange interactions. $\Pi_{\mu,\beta}$ is the momentum density conjugate to $\phi_{\mu,\beta}$.

\begin{figure}[t]
\includegraphics[height=5.0cm]{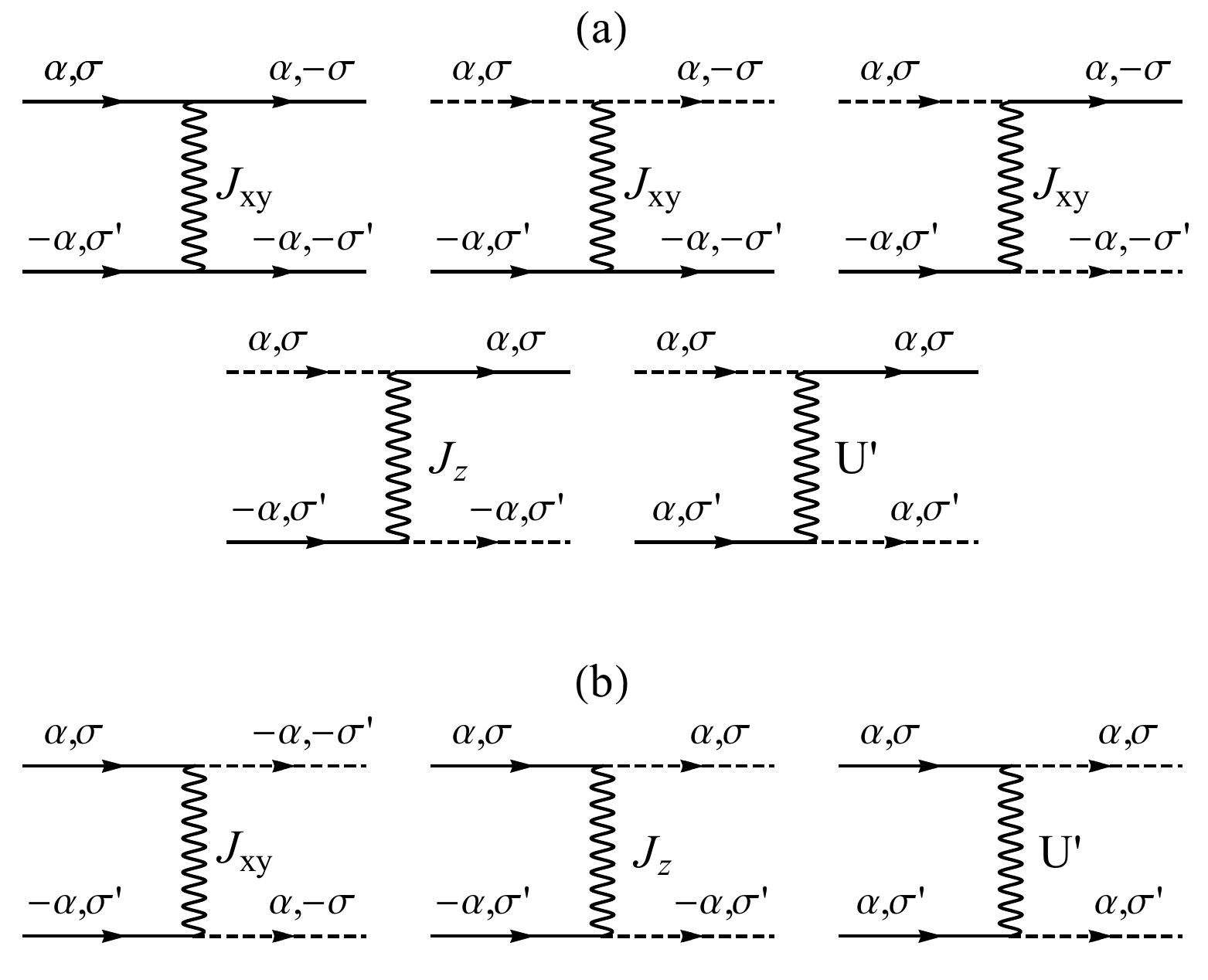}			
\caption{ The two-particle scattering processes for right-moving (solid lines) and left-moving (dashed
lines) electrons induced by Hubbard and spin exchange interactions. $\alpha$ denotes chain index and $\sigma$ denotes spin index. $-\alpha$ and $-\s$ denotes the ``flipped'' chain and spin index.(a) Momentum conserving scattering corresponding to $H_{\rm con}$; (b) Umklapp scattering corresponding to $H_{\rm umk}$. }
\label{diagram}
\end{figure}

The two-particle scattering processes (\Fig{diagram}) induced by Hubbard and spin exchange interactions introduce Sine-Gordon terms. At a generic doping level Umklapp scatterings are forbidden and the allowed momentum conserving two-particle scatterings are shown in \Fig{diagram}(a).
These scatterings give rise to the following Sine-Gordon terms:
\begin{widetext}
\bea
&&H_{\rm con} = -\frac{U^\prime}{\pi^2}\int dx~ \cos(2\phi_{+,\s})\cos(2\phi_{-,\s})+\frac{J_{\perp,z}}{4 \pi^2} \int dx  \cos(2\phi_{-,c})\left[\cos(2\phi_{+,\s}) - \cos(2\phi_{-,\s})\right]\nn \\&&+\frac{J_{\perp,xy}}{ 2\pi^2} \int dx \Big[\cos(2\phi_{+,\s})\cos(2\theta_{+,\s}) +  \cos(2\phi_{+,\s})\cos(2\theta_{-,\s}) - \cos2\phi_{-,c}\cos(2\theta_{-,\s})\Big]\nonumber \\
\eea
At half-filling, Umklapp scattering is allowed (\Fig{diagram}(b)). The additional Sine-Gordon terms are
\bea
&&H_{\rm umk} = -\frac{U'}{\pi^2} \int dx \cos(2\phi_{+,c})
\cos(2\phi_{-,c}) -\frac{J_{\perp,xy}}{2\pi^2}\int dx  \cos(2\phi_{+,c})\cos(2\theta_{-,\s})  \nonumber\\
&&+  \frac{J_{\perp,z}}{4 \pi^2} \int dx \Big[ \cos(2\phi_{+,c})\cos(2\phi_{+,\s})- \cos(2\phi_{+,c} )\cos(2\phi_{-,\s})\Big]
\eea
\end{widetext}
We are interested in FM $J_{\perp,xy},J_{\perp,z}$ and when $|J_{\perp,xy}|, |J_{\perp,z}| \gg U^\prime \gg t$.

At half-filling, the system is an insulator hence the modes are pinned by the Umklapp scattering and momentum conservation scattering, which pin $\phi_{+c}=0, \phi_{-c}=0$. Depending on the ratio between $|J_\perp|$ and $D$, different fixed points in the spin sector can be realized. According to RG analysis, in the  nearly isotropic region, $ J_{\perp,z} \approx J_{\perp,xy}< 0$ (which means $D \ll |J_\perp|$), the most relevant Sine-Gordon terms are $ J_{\perp,xy}(\cos2\phi_{+c}+\cos2\phi_{-c})\cos2\theta_{-\s}$ and $ J_{\perp,z}(\cos(2\phi_{+c}+ \cos{2\phi_{-c}})\cos2\phi_{+\s}$, which pin  $ \phi_{+\s}=0, \theta_{-\s}=\pm\frac{\pi}{2}$. As $D$ increases, the FM couplings become more anisotropic such that the most relevant Sine-Gordon terms are $ J_{\perp,xy}\cos2\phi_{+\s}\cos2\theta_{-\s}$ and $ J_{\perp,xy}(\cos2\phi_{+c}+ \cos2\phi_{-c})\cos2\theta_{-\s}$. They pin $ \phi_{+\s} = \pm\frac{\pi}{2}, \theta_{-\s}=\pm\frac{\pi}{2}$. Thus as the single-ion anisotropy $D$ increases, there is a phase transition between two distinct spin gapped phases. We identify these phases as the Haldane and ``large $D$'' phase discussed in the main text.
At the interface between these two phases, a kink occurs in $\phi_{+\s}$, which corresponds to the edge state of $S_z =\pm \frac{1}{2}$.  This is a manifestation of the topological nonequivalence of these two spin gapped phases.

Away from half-filling, the Umklapp scattering ceases to exist. As a result the total charge mode  $\phi_{+c}$ becomes gapless. However, the anti-symmetric charge mode $\phi_{-c}$ is still pinned by the momentum conservation scattering terms. This allows the two spin gapped phases to survive doping.
These are the two inequivalent topological Luttinger liquids. In particular despite both being charge gapless the interface between the doped Haldane phase and the doped large $D$ phase still exhibits the interface kink in $\phi_{+\s}$ signifying the presence of interface modes.

Because total charge mode is gapless, it is interesting to consider the superconducting pair correlation function. If we interpret the two chains as corresponding to the two atomic orbitals 1,2, only \textit{spin singlet, orbital antisymmetric and parity odd} pairing field
\bea
&&\psi_{R,1\uparrow}\psi_{L,2\downarrow} - \psi_{L,1\uparrow}\psi_{R,2\downarrow} - \psi_{R,1\downarrow}\psi_{L,2\uparrow} + \psi_{L,1\downarrow}\psi_{R,2\uparrow} \nonumber\\
&\sim& e^{-i\theta_{+c}}\left[e^{-i\phi_{-c}} \sin(\theta_{-\s} + \phi_{+\s}) + e^{i\phi_{-c}} \sin(\theta_{-\s} - \phi_{+\s})\right] \nonumber\\
\eea
shows power law decay in the doped Haldane phase. This is consistent with the DMRG results in the main text.

Interestingly, the symmetry of the pair field that exhibits power law decay in doped large-$D$ phase is quite different. The pairing channel which exhibits strongest correlation is $S_z=0$ \textit{spin triplet, orbital antisymmetric, parity even} pairing, which shows algebraic decay correlation:
\bea
&& \psi_{R,c,\uparrow}\psi_{L,d\downarrow} + \psi_{L,1\uparrow}\psi_{R,2\downarrow}
        + \psi_{R,1\downarrow}\psi_{L,2\uparrow} + \psi_{L,1\downarrow}\psi_{R,2\uparrow} \nonumber \\
&\sim& e^{-i \theta_{+c}}( e^{-i\phi_{-c}} \cos(\phi_{+\s} + \theta_{-\s}) -  e^{i\phi_{-c}} \cos(\phi_{+\s} - \theta_{-\s})). \nonumber\\
\eea
Again this is consistent with the DMRG result.

Besides cooper pair instability, gapless charge mode can also result in power-law decaying CDW correlation. However it should be noted that $2k_F$ CDW correlation decays exponentially in both doped Haldane phase and large-$D$ phase. Instead it is the  $4 k_F$ pair density wave (i.e. the CDW of cooper pairs) that exhibits power law correlation in both the doped Haldane phase and doped large-$D$ phase
\bea
O_{\rm PDW} = \psi^\dagger_{R,1,\uparrow}\psi^\dagger_{R,2,\downarrow} \psi_{L,1,\downarrow}\psi_{L,2,\uparrow} \sim e^{2i\phi_{+c}+4k_F x} e^{2\theta_{-\s}}. \nonumber\\
\eea
The decaying exponents of the PDW and SC have the relation $K_{\rm sc} K_{\rm pdw} =1$. The exponents depend on Luttinger parameter of the total charge mode: $K_{\rm sc}= 1/2 K_{+c}$, $K_{\rm pdw} = 2 K_{+c}$. When $U^\prime$ is small the superconducting correlation is dominant. As $U^\prime$  increases the pair density wave correlation becomes dominant. Irrespective of which correlation wins we have a phase fluctuating superconductor.